# Electronic Raman scattering in silicon glass

*Sergey S. Kharintsev[1,2*], Elina I. Battalova[1,2], Aleksey I. Noskov[1,2], Jovany Merham[2], Eric O. Potma[2], and Dmitry A. Fishman[2*]*

[1]Department of Optics and Nanophotonics, Institute of Physics, Kazan Federal University, Kazan 420008, Russia

[2]Department of Chemistry, University of California Irvine, Irvine, CA 92697, USA

skharint@gmail.com

dmitryf@uci.edu

ABSTRACT

The nature of enhanced photoemission in disordered and amorphous solids is an intriguing open question. A point in case is light emission in silicon, which occurs when the material is porous or nanostructured, but the effect is absent in the bulk crystalline phase, a phenomenon that is still not fully understood. In this work, we study structural photoemission in a heterogeneous cross-linked silicon glass, a material that represents an intermediate state between the amorphous and crystalline phases, characterized by a narrow distribution of structure sizes. This model system shows a clear dependence of photoemission on size and disorder across a broad range of energies. While phonon-assisted indirect optical transitions are insufficient to describe observable emissions, our experiments suggest these can be understood through electronic Raman scattering instead. This phenomenon, not commonly observed in crystalline semiconductors, is driven by structural disorder. We attribute photoemission in this disordered system to the presence of an excess electron density of states within the forbidden gap (Urbach bridge), where electrons occupy trapped states. Transitions from gap states to the conduction band are facilitated through electron-photon momentum matching, which resembles Compton scattering, but observed for visible light and driven by the enhanced momentum of a photon confined within the nanostructured domains. We interpret the light emission in structured silicon glass as resulting from electronic Raman scattering. These findings emphasize the role of photon momentum in the optical response of solids that display disorder at the nanoscale.

KEYWORDS: semiconductor glass, electronic Raman scattering, Compton scattering, Urbach bridge, photon momentum, structural optical spectroscopy

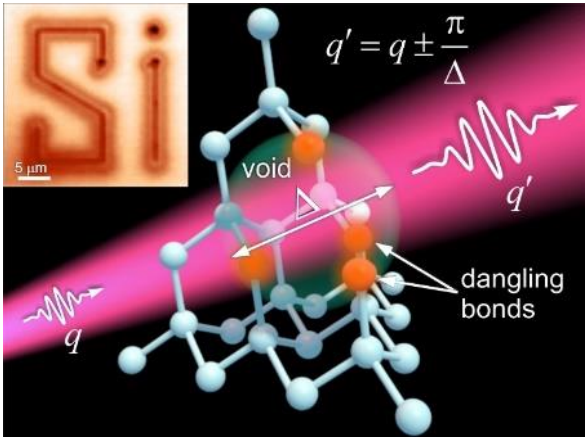

INTRODUCTION

Light absorption and emission in indirect bandgap materials is of keen interest in photovoltaics and optoelectronics.[1,2] Silicon is a case in point, as it lies at the heart of modern electronics.[3] An important challenge in photonics is to forge direct absorption and emission channels in Si throughout the visible and near-infrared range.[4,5] Tunable visible photoluminescence (PL) in porous silicon (p-Si) was first observed in the early 1990s and has opened up exciting prospects for light-emitting silicon devices.[6–10] Nonetheless, despite impressive advances ever since, the origin of PL in Si remains a subject of debate.

Quantum confinement effects, as pioneered by Canham, have been suggested as a possible explanation for the PL phenomenon.[4,11] The confinement model offers mechanisms for light emission in different forms of structured silicon, including indirect radiative recombination of excitons localized in crystalline silicon (c-Si) quantum dots,[12–14] radiative recombination of electron-hole pairs trapped at surface states of p-Si,[15] and light emission from chemical impurities anchored to the structural defects and asperities of p-Si.[7,16–18] All aforementioned emission mechanisms indicate an increase of the local electronic density of states (e-DOS) at nano- and sub-nano structures, which include defects, dangling bonds and chemical moieties.[19,20] Unfortunately, the quantum confinement model falls short in elucidating the anomalous PL redshift observed in Si nanocrystals.[21,22]

Our study delves into the photoemission of a heterogeneous cross-linked Si glass produced by continuous wave (cw) laser annealing of amorphous silicon (a-Si). The glass comprises crystalline Si nanoclusters cross-linked to an amorphous matrix. Upon illumination with cw radiation, the glass emits two distinct bands, previously denoted as low-energy PL (band S) and high-energy PL (band F),[6] whose origins have remained unclear. Here, we suggest that both emission bands originate from electronic Raman scattering (ERS).[23,24]

One of the first observations of ERS in semiconductors has been linked to direct optical transitions between a light-hole band and a heavy-hole band,[25] an effect distinct from vibrational Raman scattering (VRS) where the initial and final electronic states remain identical. We attribute the observed emission to ERS transitions from states within the bandgap, which, in the context of the Mott-Davis model for disordered semiconductors,[26,27] are expected to be present in the cross-linked silicon glass studied here. Because such Urbach states constitute trapped states for the electron, transitions in close proximity to the conduction band require a source of momentum. Although lattice phonons have been proposed to supply the necessary momentum, higher-energy transitions from deep trapped states would require the involvement of multiple phonons,[28] thus severely limiting the probability of such transitions. We propose that transitions from these states are enabled instead by electron-photon momentum matching, a result of quantum confinement,[29] particularly in a disordered medium. The concept of expanded near-field photon momentum has previously been proposed as a plausible explanation for enhanced interband two-photon excitation[30] as well as intra-band transitions in gold nanostructures.[31] More recently, momentum expansion in plasmonic two-dimensional systems has been discussed, focusing on multipolar, spin-flip, and multi-quanta emission processes.[32] Here we suggest that the existence of near-field photon modes with expanded momenta in a Si glass provides the necessary momentum to facilitate ERS transitions from trapped states in the bandgap. This effect is similar to Compton scattering, though with visible light photons.

In the context of disordered semiconductors, we revisit an idea proposed and developed by Mott and Davis[27,33] and use it to explain the PL phenomenon. The model is based on the concept of dangling bonds at vacancies, divacancies or nano-voids. It introduces an excess electronic density of states within the forbidden gap,[34] forming an Urbach bridge of electronic states across the semiconductor bandgap. This bridge enables new electronic Raman transitions, including low-energy intra-band transitions near the Fermi level ($l$-ERS), higher energy band from inter-band transitions in the extended tail near the conduction band ($h$-ERS) and its heavy tail indicating optical transitions from deep states in the Urbach bridge.

These emissions exhibit a strong correlation with structural size, a relationship that can be explained through the concept of electron-photon momentum matching. Given the direct link between optical signals and the formation of cross-linked semiconductor glasses, the findings presented here pave the way for expanding conventional optical spectroscopy for both chemical (energy) and structural (momentum) studies of disordered solids.

RESULTS AND DISCUSSION

First, we offer a strategy to produce samples where both crystallinity and structure size are known and formed in a controlled fashion. We consider a bottom-to-top approach, i.e. from a disordered state to a more ordered material state, by which amorphous silicon (a-Si) transforms in part into crystalline silicon (c-Si) through light-assisted thermal impact. This approach allows efficient photon absorption in an amorphous matrix, followed by light emission at c-Si nanocrystal inclusions. To accomplish this, we deposit a 300 nm thick a-Si film on glass using chemical vapor deposition (see *Methods*). The film is then subjected to a tightly focused cw laser beam, which is scanned to write an array of straight lines at different scanning speeds (Figure 1a). The high light absorption ($\alpha = 83\,870$ cm$^{-1}$ at 633 nm[35]) and low thermal conductivity ($\kappa = 1.7 - 2.2$ W/mK [36]) of a-Si give rise to local heating. In areas where light-induced temperature changes do not exceed 500 °C, this procedure results in the sintering of amorphous structures forming a homogeneous cross-linked glass (Figure 1b).[37,38] In areas where temperature exceeds 500°C, an amorphous-to-crystalline phase transition occurs, and a heterogeneous cross-linked semiconductor glass is formed. The resulting film, 'foamed' by light, represents a heterogeneous disordered matrix in which electronic, optical and thermal properties vary on the nanometer scale.

In our experiment, the local degree of crystallinity is determined by the speed of the scanning laser beam with an intensity of 2.5 MW/cm$^2$, which was set to 0.5, 1, 2, 4 or 8 μm/s.

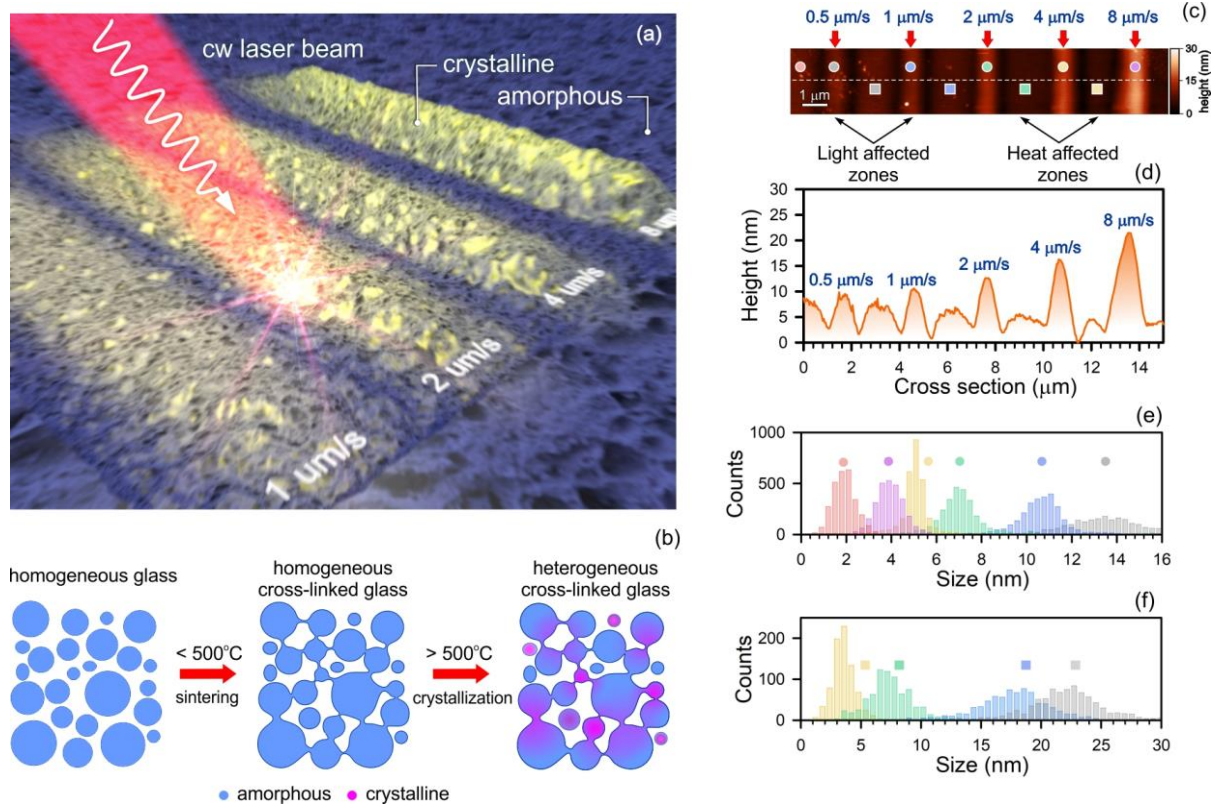

**Figure 1.** (a) An artistic representation of light-assisted formation of cross-linked semiconductor glass stripes on a-Si film. (b) Schematic representation of temperature dependent formation of homogeneous cross-linked glass (through sintering) and heterogeneous cross-linked glass (through crystallization). (c) AFM topography and (d) cross-section of an array series. By varying exposure time (writing speed), one can form areas with different degree of crystallinity and a narrow distribution of sizes within light affected zones (e) and heat-only affected zones (f).

An AFM topography map in Figure 1c shows that an array of stripes can be easily formed, where their height increases proportionally with the scanning rate, reaching a few tens of nanometers. For intensities above 3 MW/cm$^2$, we observe the formation of bubbles on the surface of the film (see *Supplementary Information Part I*, Figure SF1). These bubbles are prone to bursting, leading to the formation of significant protrusion areas that were used to assess the thickness of the initial a-Si film. Within the formed glass surface, we identify two distinct areas. The first is the light-affected zone (LAZ), which represents the area directly exposed to the laser radiation. The second is the heat-only affected zone (HAZ), which encompasses the portion that remained unexposed to the laser and was solely influenced by diffusion of heat.

Figure 1d shows structural growth within the LAZ. Changes in the film topology are caused by more a compact arrangement of atomic Si upon crystallization, leading to relaxation of intrinsic local stress to minimize the Gibbs energy. Here, we come across a counterintuitive observation: extended exposure times result in inferior crystallization and smaller structural formation. These can be explained by the larger thermal conductivity of c-Si ($\kappa_{c\text{-Si}} = 147$ W/mK) compared to that of a-Si ($\kappa_{a\text{-Si}} = 1.7 - 2.2$ W/mK). For this process, a negative feedback loop is initiated, causing crystallization to cease as a result of efficient heat transfer. The local temperature may drop below the threshold of 500°C, while the film morphology continues to change due to sintering (Figure 1e).

The dissipation of heat beyond the LAZ induces changes in morphology and structure formation in the HAZ. Figure 1f illustrates a broader distribution of surface roughness in HAZ compared to LAZ. It can be seen that HAZ differs from LAZ by the lack of a crystalline (c-Si) phase. The degree of crystallinity in each zone is monitored using vibrational Raman scattering (VRS).[24] While LAZ is clearly rich of c-Si, the temperature in HAZ does not exceed the required threshold. Meanwhile, crystallization in HAZ can be triggered by pressure and local stress[39] at amorphous/crystalline interfaces, an effect that is observed in our experiments near the zone boundaries (Figure 1f). In summary, laser writing yields a narrow distribution of nanostructures sizes within both LAZ and HAZ, as estimated from the surface roughness. This model system proves invaluable for exploring structure- and phase-dependent photoemission.

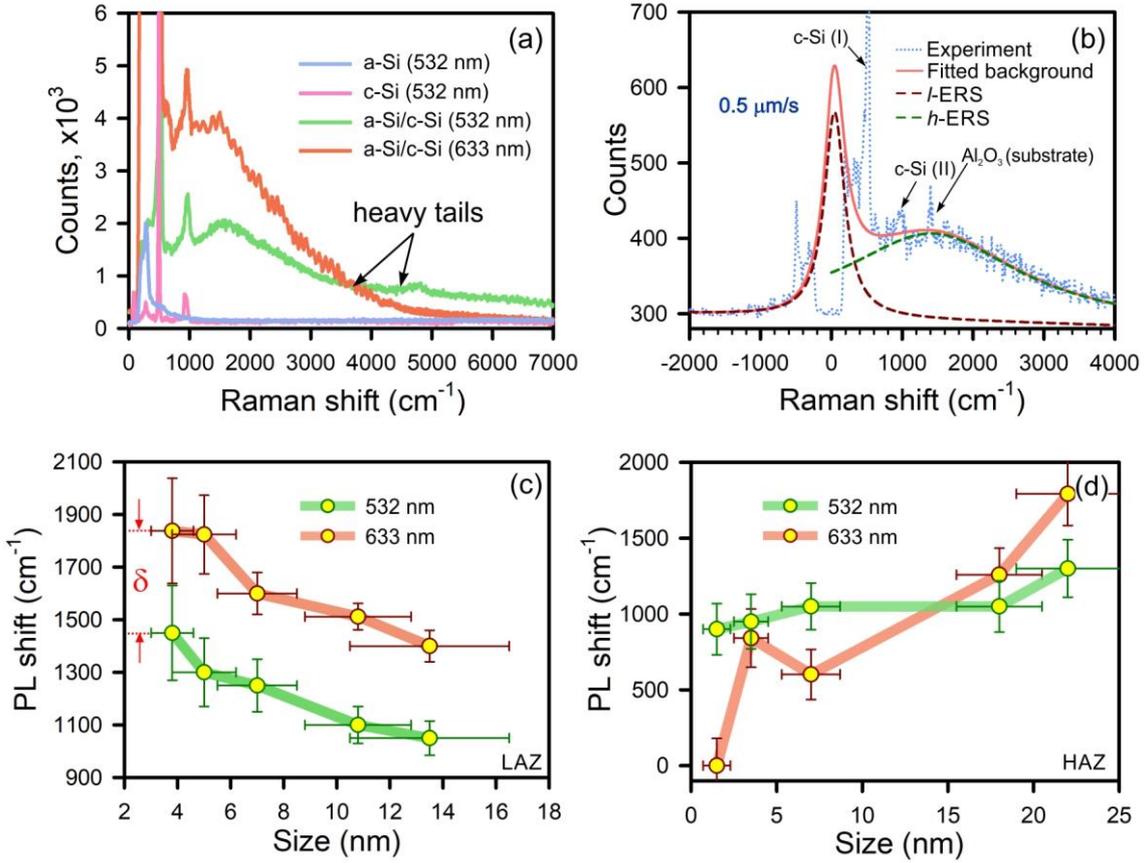

**Figure 2.** (a) Raman spectra of a-Si (light blue), c-Si (pink) using 532 nm excitation and silicon glass at LAZ (0.5 μm/s) using 532 nm (green) and 633 nm (orange) excitation wavelengths, (b) Decomposition of Raman spectrum at LAZ written at 0.5 μm/s. (c) and (d) The *h*-ERS shift vs size for LAZ and HAZ using 532 nm and 633 nm excitation wavelengths.

The Raman spectrum of LAZ, obtained at a position where a writing speed of 0.5 μm/s was applied, is shown in Figure 2a. A rich emission spectrum covering a broad energy range is observed, where the presence of a crystalline phase can be inferred from the lines at 521 cm$^{-1}$ and 960 cm$^{-1}$, attributed to the first- and second-order (optical) phonon modes of c-Si. The broad emission band, extending well beyond 5000 cm$^{-1}$, peaks near 1700 cm$^{-1}$ for either the 532 nm or 633 nm excitation wavelength. The overall invariance of the emission maximum with excitation color rules out a dominant role for radiative recombination originating from thermalized electron populations in the conduction band and holes in the valence band. In addition, the observation of the emission at large Stokes shifts (>3000 cm$^{-1}$), which we refer to as the *heavy tail*, cannot be accounted for in the context of thermalization of electrons and holes (Supplementary Information Part II). Phonons are required for the thermalization process, yet the vibrational density of states (v-DOS) of the phonon bath is expected to be reduced for nanoscale structures, which can prolong the duration of thermalization by an order of magnitude.[22,40] A lower v-DOS becomes negligible for structures smaller than a few nm, so that the probability of emitting or absorbing phonons, needed for electron thermalization and indirect phonon-assisted transitions, is significantly reduced. Moreover, quantum confinement should raise the bottom edge of the conduction band at the $\Gamma$-X point of the Brillouin zone by 1 eV (Figure SF3). This means that the 633 nm photon carries insufficient energy to induce indirect inter-band transitions in such sub-nanometer structures. These considerations are at odds with a model that relies on emission from radiative electron-hole recombination, extensively discussed in *Supplementary Information Part II*, but comply with the ERS model for emission.

Figure 2b displays the full LAZ emission spectrum, showing spectral features on both the Stokes and the anti-Stokes sides. In the following, we ignore the optical phonon signatures and focus solely on the broader spectral features. Using a regularized least squares method, we decompose the spectral response into two distinct bands denoted as *l*-ERS (low-energy ERS near Fermi level, red dotted line, Figure 2b) and *h*-ERS (high energy shifted ERS band, green doted line, Figure 2b). We note that both *l*-ERS and *h*-ERS are absent in bulk c-Si – they only appear after the light-induced structuring in the a-Si material.

Figures 2c presents the central energy shift of *h*-ERS as a function of structural size using an excitation wavelength of either 633 nm or 532 nm (depicted by the blue solid curves). These plots reveal a clear trend: a pronounced redshift of *h*-ERS as the structure size is decreased. Below, we argue that this observation provides evidence for ERS from trapped states in nanocrystalline inclusions in the a-Si matrix. We observe an opposite trend for the size-dependence of the *h*-ERS energy shift in HAZ. Figure 2d shows a growth of the energy shift as the structure size is increased, which is observed for both excitation wavelengths. This latter phenomenon can be explained in the context of an amorphous semiconductor, which constitutes the only component in HAZ. In amorphous Si the energy band structure is smeared compared to the crystalline phase, associated with a decreasing bandgap when the structure size increases (Supplementary Figure SF4).[41,42] In the case of unperturbed a-Si with a ~2 nm surface roughness, the *h*-ERS fully overlaps with *l*-ERS (see *Supplementary Information Part IV* and Figure SF4). As the a-Si film is subjected to heat, its roughness increases due to the sintering process, leading to a redshift in the *h*-ERS, a result of size-dependent band smearing.

To explain the size dependent emission energy shift in LAZ, we propose an alternative explanation for the emission origin in cross-linked semiconductor Si glasses that is based on electronic Raman scattering enabled by electron-photon momentum matching, as depicted in Figure 3a.[23,24] This model considers that, in disordered semiconductors, localized electron states form an excess electronic density of states (e-DOS) across the forbidden gap, extending from the band edges down to Fermi level (Figure 3b and inset Figure 3a), as was predicted by Mott et al.[27,33] Molecular dynamic simulations reveal that defect-induced Hellman-Feynman forces govern the forbidden gap,[43] linking the upper valence edge band and the bottom conductance edge band. We term this electronic continuum "the Urbach bridge", that is conceptually illustrated in Figure 3b and the inset of Figure 3a. *The size-dependent closing of the bandgap transforms an amorphous semiconductor into a quasi-metal and enable the observation of indirect and direct optical transitions*.

Within this model, the observed *l*-ERS emission can now be attributed to optical transitions in the vicinity of the Fermi level. Here, we establish a connection between the observed *l*-ERS and emission effects that are associated with photon momentum, phenomena previously reported in rough metals,[30,31] disordered semiconductors,[44] and high entropy oxides.[45] Similar to metals, the *l*-ERS peak remains centered at the Rayleigh line for varying sizes of the crystallites, yet a clear correlation exists between its linewidth ($\Gamma$) and structure size (Figure 3a). An empirical connection between the disorder-driven Raman linewidth ($\Gamma$) and Urbach energy ($E_U$), i.e. $\Gamma \sim E_U$ has been previously established.[46,47] Our data demonstrates that the *l*-ERS linewidth (Urbach energy) exhibits an exponential growth that is inversely proportional to the structure size (Figure 3 a). At 4 nm, the *l*-ERS linewidth reaches the value of 43 meV (350 cm$^{-1}$), while for larger structures, it asymptotically approaches the thermal energy $kT \approx 23\,\text{meV}$ (200 cm$^{-1}$). It is crucial to note that the *l*-ERS peak diminishes as the cross-linked glass undergoes excessive or full crystallization.

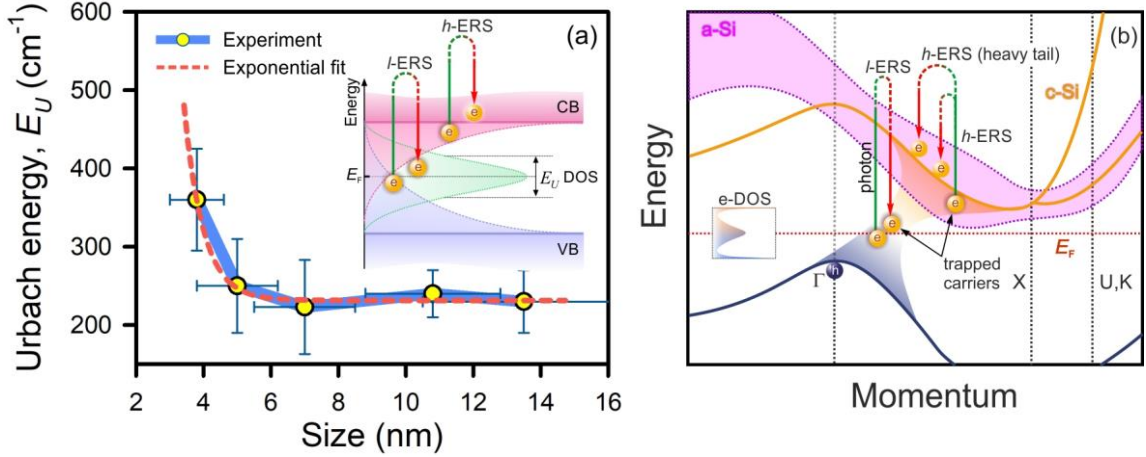

**Figure 3.** (a) The *l*-ERS linewidth at LAZ (Urbach energy) as a function of structure size. The inset shows a schematic representation of Urbach bridge following the concept idea by Mott and Davis.[27,33] (b) Conceptual visual representation of direct and indirect optical transitions for understanding the origins of *l*-ERS and *h*-ERS.

Applying a similar rationale, the *h*-ERS and its energy redshift is explained as an electronic Raman process linked to optical transition from the Urbach bridge to the conduction band (Figure 3b). To extract an electron from the trapped state within the bridge in the mobility gap, a change in electron momentum is necessary.[48] While previous studies by Zhang and Drabold[49] have explored the use of phonons to assist in this transition, the authors acknowledge that the energy of a single phonon would not suffice for enabling the transition deep from the mobility gap as observed by the emission's heavy tail (Figure 2f, >4000 cm$^{-1}$). Alternatively, both energy and momentum required for such transitions can be supplied by the photon confined within the nanostructure. The photon's large energy and expanded momentum enable indirect optical transitions from deeper states within the Urbach bridge to the conduction band. This notion is further supported by the observed difference in the heavy tail intensity when using excitation wavelengths of 532 nm or 633 nm (Figure 2a). While the center of mass of the *h*-ERS spectrum remains similar for both cases, the heavy tail appears slightly more intense with the use of 532 nm excitation. While the width of the phonon's *k*-spectrum in both cases is fully defined by spatial nanostructure size,[49] confinement of a 532 nm photon should yield a higher optical density of states compared to a confined 633 nm at the same structure. Hence, for a higher amplitude *k*-spectrum at 532 nm, the wings of the spectral distribution contribute more significantly to the transitions where larger momenta are required (Figure SF5).

To facilitate a careful examination of the data within the context of the proposed Urbach bridge concept, Figure 4 shows Raman spectral maps summarizing the overall trends and correlations observed for two distinct types of structures. The first type, extensively discussed earlier, consists of 1D arrays (Figure 1, Figure 4a-e). The second one comprises a 2D network of cross-linked semiconductor glass (Figure 4f-j). It has been fabricated on a similar a-Si film through scanning the laser beam and forming a network of crossed protrusions with a gradually adjusted pitch, as imaged in AFM topography (Figure SF7a). The spectral maps acquired from these structures not only establish clear correlations, but affirm a direct connection between structure size, degree of crystallinity and their corresponding spectral responses.

Figures 4a-b and 4f-g show spectral maps of *vibrational (VRS) Raman scattering* originating from a-Si and c-Si. An inversion of the VRS maps for a-Si and c-Si is evident. Following the discussion above, the degree of crystallinity is intricately tied to the writing speed, driven by a substantial difference in thermal conductivity between a-Si and c-Si.

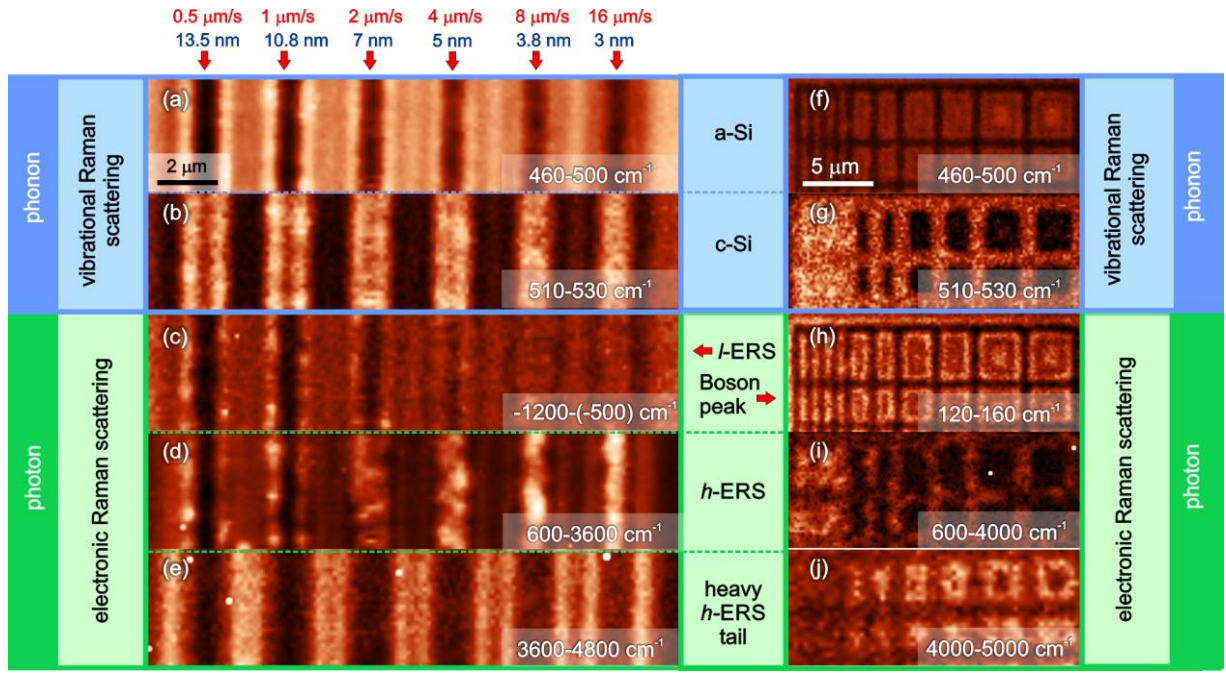

**Figure 4.** An 1D array of straight lines fabricated on a-Si film using different laser scanning speeds: VRS maps (a-Si (a) and c-Si (b)) and ERS maps (*l*-ERS (c) and (d),(e) *h*-ERS). A light-structured cross-linked network on a-Si film at different pitch: VRS maps (a-Si (f), c-Si (g) and a boson peak (h)) and ERS maps ((i),(j) *h*-ERS).

Hence, in the case of 1D array, the crystallization in the LAZ and, consequently, the c-Si VRS are depressed when a slower writing speed (0.5 μm/s) is used. At the same time, the boundaries between LAZ and HAZ contain substantial c-Si nanostructures, as confirmed by AFM (Figure 1d), causing these regions to light up in this spectral range. With an increase in writing speed, crystallization proceeds more uniformly, leading to a decrease in the VRS intensity at HAZ boundaries.

ERS maps (*l*-ERS and *h*-ERS) for both types of samples follow the spatial distribution of structural sizes and degree of crystallinity across their surfaces (Figure 4c-e and 4i-j). To visualize *l*-ERS for the 1D array, we plot the integrated signal at the anti-Stokes wing in the -1200-(-500) cm$^{-1}$ spectral range (Figure 4c), avoiding contributions from other spectral features, particularly the Boson peak. In line with the discussion on data in Figure 3a, the trend shows an increase in the integrated signal (*l*-ERS width, Figure 4c) with a decrease in structure size.

Furthermore, a Boson peak map at 140 cm$^{-1}$ for the 2D network is plotted in Figure 4h (also see Figure SF7e). The Boson peak emerges in disordered systems due to the hybridization of plane-wave bulk and confined phonons.[50,51] As can be seen on the spectral map, this peak intensifies with crystallization near the LAZ/HAZ interface due to an increase in v-DOS of bulk phonons - an observation consistent with the VRS maps (Figure 4a-b and 4f-g) discussed earlier. The Boson peak, however, proves insensitive to thermal impact, as observed in HAZs (Fig. SF6b), although it shows a slight degradation compared to that of a-Si. This phenomenon may be attributed to thermal-induced amorphization, resulting in increased roughness and stress relaxation.

The *h*-ERS maps, generated by integrating the signals above 600 cm$^{-1}$, are shown in Figures 4d-e (1D array) and 4i-j (2D network). In the case of the 1D array, the *h*-ERS intensity in LAZ increases with writing speed. *Supplementary Information Part VI* provides additional maps generated by integration over a narrower spectral range (Figure SF5). These results corroborate earlier observations detailed in Figure 2c and 2d - the *h*-ERS energy shift is inversely proportional to the structure size. Following surface topography and the spatial distribution of c-Si nanocrystals, the *h*-ERS extends further away from LAZ and towards HAZ. The *h*-ERS maps of 2D network exhibit a similar trend (Figure 4i-j). Here, an interesting observation is noted: multiple passes of the laser beam through closely packed spatial locations seem to influence morphology (or size range), but not the level of crystallinity. This conclusion naturally follows from the direct comparison between the c-Si VRS map (Figure 4g) and the Si glass ERS map (Figure 4i). In areas where multiple laser passes create a smaller pitch (left side of the map, Figure 4i), the ERS signal intensifies. The closely packed pitch 'foams' the film, causing it to become more heterogeneous by blending amorphous and crystalline phases.

The spectral maps of the ERS heavy tail are presented in Figure 4e and 4j. Clearly, the *h*-ERS extends beyond the LAZ and the highest energy-shifted ERS signals are observed well into the HAZ. We note that the HAZ is not expected to contain large, fully developed c-Si structures, as preliminary confirmed by c-Si VRS (Figure 4b). However, these areas may host crystal embryo's - sub-to-near nm Si crystallization nuclei - which can be present near the LAZ/HAZ interface and extend into the HAZ due to the pressure gradient between the zones.[39] These intriguing sub-areas can be referred to as pressure-affected zones (PAZ). A photon confined at such crystalline embryos carries both energy and a significantly large momentum, which can be transferred to an electron, facilitating the transition from deep trapped states within the forbidden gap to the conduction band (Figure 3b). This process gives rise to a substantial energy-shifted ERS heavy tail, extending it spectrally well above 4000 cm$^{-1}$ and observable up to 7000 cm$^{-1}$ (Figure SF11). Such energy shifts cannot be explained using conventional electron-phonon interactions.[29,43,52] To demonstrate the effect of pressure driven Si-formations, we conducted a series of experiments using an AFM cantilever (see *Supplementary Information Part VIII*). In these experiments, the tip was used in contact tapping mode to create pressure points on an a-Si film. The regions subjected to such pressure exhibit a significant increase in the heavy tail of *h*-ERS, while spectra show no presence of c-Si (Figure SF9a, orange spot and corresponding spectrum on Figure SF9c). Furthermore, a correlation between Boson peak and heavy tail ERS is clearly observed, specifically the dark inner frames in Figures 4h and the bright frames in Figure 4j (Supplementary Figures SF7e and SF7f). Remarkably, these regions were not directly impacted by laser illumination. Nonetheless, the observed correlation suggests the presence of crystalline embryos formed by local internal stress when the film topology undergoes changes due to laser writing.

CONCLUSION

This work presents a detailed study of light emission in a heterogeneous cross-linked glass composed of c-Si nanocrystals embedded in an amorphous silicon matrix. With such a model sample at hand, we attempt to address a few outstanding questions regarding the origin of PL and discuss the emission spectrum in relation to crystal phase, size and photon excitation energy. Our findings challenge the conventional notions of phonon-assisted fluorescence in quantum confined systems. To interpret our observations, we revisit the concept proposed in 1967 by Mott and Davis.[26] They suggested that a continuum of energy states forms in the forbidden gap once disorder and crystallinity are presented at the nanoscale. This extended e-DOS establishes a quasi-continuous connection between the conduction and valence bands, referred to here as an Urbach bridge. We propose that intra- and inter-band optical transitions within Urbach bridge underlie the observed emission effect.

The observation and spectral analysis of the *l*-ERS leads us to suggest electronic Raman scattering as the dominant mechanism of emission, a phenomenon previously observed and explained in the context of extended photon momentum on nanoscale asperities of metal surfaces[30] and even individual gold particles.[31] The electronic Raman nature of the emission is further supported by a substantial *h*-ERS redshift, its correlation with size, and lack of dependence on excitation photon energy.

In our proposed model, both emission features originate from the Urbach bridge, a conclusion supported by the synchronous appearance and evolution of the signals upon mixing amorphous and crystalline phases. Both features disappear when the system fully transforms to bulk c-Si. The *l*-ERS on the Urbach bridge is associated with ERS transitions near the Fermi level. We argue that confined photons with expanded momentum likely exist in a nanostructured a-Si/c-Si matrix. Inter-band optical transitions from the Urbach bridge to the conduction band are responsible for *h*-ERS. This insight helps explain its heavy tail (>4000 cm$^{-1}$ energy shift, see Figure SF10 and Figure SF11) originating from deep states in the forbidden gap and to the conduction band. These transitions require the smallest structural confinement and large photon momentum available at embryonic c-Si sites. The size dependence of the *l*-ERS and *h*-ERS underscores their potential utility as spectroscopic probes for quantifying the structural disorder in vitreous semiconductors and empower optical spectroscopy to enable structural analysis of disordered solids.

Lastly, we would like to highlight an apparent similarity between the ERS and Compton scattering processes. The Compton effect occurs when a propagating photon with a relatively large momentum, i.e. X-ray, scatters upon an electron.[53] Similarly, the ERS phenomenon discussed here is driven by a confined visible photon with enhanced momentum, scattering on an electron in a trapped state within the forbidden energy gap. The matching of electron and photon momenta enables these light-matter interactions that are otherwise forbidden. While one of the conventional and efficient methods to confine light involves utilizing plasmon resonance, its efficacy diminishes when the structure size is less than 5 nm. Meanwhile, structural singularities, including but not limited to vacancies, dangling bonds, crystalline embryo's, induce an electrostatic lightning rod effect,[54] significantly expanding photon momentum and surpassing the plasmonic effect.[55,56]

**ACKNOWLEDGEMENT**

D.A.F. and S.S.Kh. would like to thank Yulia Davydova and Natalia Bratkova for help and support. Authors thank Prof. Alexander Fishman, Prof. Sasha Chernyshev and Prof. Maxx Arguilla for the fruitful discussions. S.S.Kh. thanks RSF grant No. 19-12-00066-P for experimental studies of cross-linked heterogeneous silicon glass. D.A.F and E.O.P. are thankful to Chan-Zuckerberg Initiative and grant 2023-321174 (5022) GB-1585590 and EOP thanks the National Science Foundation, grant CMMI-1905582 for developing a theoretical model of light emission in silicon glass. All authors acknowledge a technical support from NT-MDT BV (The Netherlands).